\documentclass[twocolumn,amsmath,amssymb,aps,pra,showkeys]{revtex4}%%{revtex4-1}
\usepackage{graphics,epsfig}
\usepackage[]{graphicx}
\usepackage{latexsym}
\usepackage{amsmath, amsthm, amsfonts, amssymb}
\usepackage{verbatim}
\usepackage{graphicx}% Include figure files
\usepackage{dcolumn}% Align table columns on decimal point
\usepackage{bm}

\begin{document}

\preprint{APS/123-QED}
\newcommand{\be}{\begin{equation}}
\newcommand{\ee}{\end{equation}}
\newcommand{\ben}{\begin{eqnarray}}
\newcommand{\een}{\end{eqnarray}}
\newcommand{\n}{\nonumber  }
\newcommand{\nn}{\nonumber \\ }
\newcommand{\p}{\partial}
\newcommand{\nd}{\noindent}
\newtheorem{theo}{Theorem}[section]
\newtheorem{definition}[theo]{Definition}
\newtheorem{lem}[theo]{Lemma}
\newtheorem{prop}[theo]{Proposition}
\newtheorem{coro}[theo]{Corollary}
\newtheorem{exam}[theo]{Example}
\newtheorem{rema}[theo]{Remark}
\newtheorem{example}[theo]{Example}
\newtheorem{principle}[theo]{Principle}
\newcommand{\ninv}{\mathord{\sim}} %involutive negation
\newtheorem{axiom}[theo]{Axiom}
%\numberwithin{equation}{subsection}

\title{Quantal effects and  MaxEnt}

\author{F. Holik}
\affiliation{Departamento de Matem\'{a}tica - Ciclo B\'{a}sico Com\'{u}n\\
Universidad de Buenos Aires - Pabell\'{o}n III, Ciudad
Universitaria \\ Buenos Aires, Argentina\\% with \\
}%
\affiliation{
 Postdoctoral Fellow of CONICET\\
}%
\author{A. Plastino}
\affiliation{%
 National University La Plata
\& CONICET IFLP-CCT, C.C. 727 - 1900 La Plata, Argentina \\  Universitat de les Illes Balears and IFISC-CSIC \\
 07122 Palma de Mallorca, Spain \\
Instituto Carlos I de Fisica Teorica y Computacional and
Departamento de Fisica Atomica, Molecular y Nuclear \\ Universidad
de Granada, Granada, Spain
}%

\date{\today}% It is always \today, today,

\begin{abstract}
\noindent Convex operational models (COMs) are considered  as
great
 extrapolations to larger settings of any statistical theory. In this article we
 generalize the maximum entropy principle (MaxEnt) of Jaynes' to
any COM. After expressing Max-Ent in a geometrical and latttice
theoretical setting, we are able to cast it for any COM. This
scope-amplification opens the door to a new systematization of the
principle and sheds light into its geometrical structure.
\begin{description}
\item[PACS numbers]
\textbf{03.65.Ud}
\end{description}
\end{abstract}

\pacs{Valid PACS appear here}% PACS, the Physics and Astronomy
\keywords{entanglement-quantum separability-convex
sets}%Use showkeys class option if keyword
\maketitle

\bibliography{pom}

\section{Introduction}

\nd The notion of using a small set of relevant expectation values
so as to describe the main properties of physical systems may be
considered the leit-motiv of statistical mechanics \cite{brillu}.
Developments based upon Jaynes' maximum entropy principle (MaxEnt)
constitute a pillar of our present understanding of the discipline
\cite{jaynes,katz}. This type of ideas has also been fruitfuly
invoked for obtaining the probability distribution associated to
pure quantum states via MaxEnt (see for instance \cite{pp} and
references therein). Indeed, MaxEnt constitutes a very important
physical viewpoint, on the one hand, and  powerful technique on the
other one. This is true not only for physics, chemistry, astronomy,
and engineering but for a host of other disciplines as well
\cite{miloni}. Thus, MaxEnt extensions that amplify its range of
applicability should be regarded as significant for all branches of
science.

\nd In this communication we wish to introduce the powerful Convex
Operational Models (COM) approach
\cite{Barnum-Wilce-2006,Barnum-Wilce-2009,Barnum-Wilce-2010} into
the MaxEnt domain. COM is, in turn, intimately linked to convex sets
of probability measures
\cite{Barnum-Wilce-2006,Barnum-Wilce-2009,Barnum-Wilce-2010,Beltrametti.Varadarajan-2000,Gudder-StatisticalMethods,Cattaneo-Gudder-1999}.
Among the several extant approaches to the study of convex sets of
probability measures, the above cited references deal with the
COM-technique, in which physical states (understood as probability
measures) and their convex structure play a key role, while other
related quantities emerge in rather natural fashion. There exist
generalizations of quantum mechanics,  including non-linear
versions, that are axiomatized using the convex structure of the set
of states (see \cite{MielnikGQS}, \cite{MielnikTF}, and
\cite{MielnikGQM}). Using convex sets one treats in geometrical
fashion the statistical theory of systems, and can also include
quantum and classical mechanics (indeed several other theories as
well). Elementary (sharp) tests in quantum mechanics are represented
by projection operators that form the well known von Newmann's
lattice, an orthomodular one \cite{BvN}. Vis the Born's rule any
projection operator defines probabilities. These are linked to
measures over the von Newmann's lattice: using Gleason's theorem, it
is possible to link in a bijective way density matrixes and
non-kolmogorovian probability measures. But probability measures may
also arise in quantum mechanics by means of positive operator valued
measures (POVMs), which are also known as generalized measures.
POVMs are also generalized easily to any COM. We will review these
subjects in Section \ref{s:COMpreliminaries}.

\nd We will express here Jaynes's MaxEnt appproach to a lattice
theoretical form, and using this, extend it to arbitrary COM, which
entails a considerable enlargment of its scope. The extension is
done in such a way that includes conditions on POVMs. This is done
in Sections \ref{s:NewWay} and \ref{s:Generalization}. Such
generalization will undoubtedly be of interest to thousands of
MaxEnt practitioners. Since COMs constitute important extrapolations
to larger settings of any statistical theory, casting MaxEnt for any
COM, after expressing Jaynes' principle in a geometrical and lattice
theoretical scenario, will hopefully open the door to a new
systematization of the principle while shedding light into its
geometrical structure.

\section{Maximum Entropy Principle (MaxEnt)}\label{s:convex set of states}

\nd Statistical mechanics and thereby thermodynamics can be
formulated on the basis of information theory if the  density
distribution $\rho(x)$ is obtained by recourse to MaxEnt
\cite{jaynes,katz}. Jaynes' stance asserts that assuming that your
prior knowledge about the system is given by the values of $n$
expectation values of physical quantities $R_j$, i.e., $\langle R_1
\rangle,\ldots,\langle R_n \rangle$, then the most unbiased
probability distribution $\rho(x)$  is uniquely fixed by extremizing
Shannon's logarithmic entropy $S$ subject to the $n$ constraint

\begin{eqnarray}\label{e:conditionsmean} \langle
R_{i}\rangle=r_{i};\,\, {\rm for \,\,all\,\,} i.
\end{eqnarray}

This brings into the game $n$ Lagrange multipliers $\lambda_i$.
%It is usual in appealing to information theory tools (like $S$) to
%regard the accompanying PDFs as being dimensionless quantities.
In the process of employing the MaxEnt procedure one discovers that
the information quantifier $S$ can be identified with the
equilibrium entropy of thermodynamics if our prior knowledge
$\langle R_1 \rangle,\ldots, \langle R_n \rangle$ refers to
extensive quantities \cite{jaynes}. $S(maximal)$, once determined,
yields complete thermodynamical information with respect to the
system of interest \cite{jaynes}. The MaxEnt probability
distribution function (PDF), associated to Boltzmann-Gibbs-Shannon's
logarithmic entropy $S$, is given by \cite{jaynes,katz}

\begin{equation} \label{z1}
\rho_{max}=\exp{[(-\lambda_{0}\mathbf{1}-\lambda_{1}R_{1}-\cdots-\lambda_{n}R_{n})]},
\end{equation}

\noindent where the $\lambda$'s are Lagrange multipliers
guaranteeing that

\begin{equation}  \label{z2}
r_{i}=-\frac{\partial}{\partial\lambda_{i}}\ln Z,
\end{equation}

\noindent while the partition function reads

\begin{equation}  \label{z3}
Z(\lambda_{1}\cdots\lambda_{n})=\mbox{tr}[\exp^{-\lambda_{1}R_{1}-\cdots-\lambda_{n}R_{n}}],
\end{equation}
\noindent and the  normalization condition

\begin{equation}  \label{z4}
\lambda_{0}=\ln Z.
\end{equation}
Such simple-looking algorithm constitutes one of the most powerful
ones in physics' arsenal. In a quantum setting, of course, the $R$'s
are operators on a Hilbert space $\mathcal{H}$ while $\rho$ is a
density matrix (operator).

\section{COM-preliminaries}\label{s:COMpreliminaries}

\nd We recapitulate here essential aspects of convex operational
models. ${\mathcal{P}}({\mathcal{H}})$ will denote the set of all
closed subspaces of $\mathcal{H}$, which are in a one to one
correspondence with the projection operators. Because of the one to
one link, one usually employs the notions of ``closed subspace'' and
``projector'' in interchangeable fashion. The bounded operators on
$\mathcal{H}$ will be denoted by $\mathcal{B}(\mathcal{H})$. We
begin the present task with reference to classical probabilities.
Given a set $\Omega$, let us consider a $\sigma$-algebra
$\Sigma\subseteq\mathcal{P}(\Omega)$. Then, a probability measure
will be given by a function

\begin{equation}
\mu:\Sigma\rightarrow[0,1],
\end{equation}
\noindent which satisfies the well known axioms of Kolmogorov
\cite{kolmog}.
 \nd As for quantum probabilities we remind
that, in the standard formulation of quantum mechanics, states may
be defined as functions of the form \cite{mikloredeilibro}

\begin{equation}\label{e:nonkolmogorov}
s:\mathcal{P}(\mathcal{H})\rightarrow [0;1],
\end{equation}
\noindent such that:

\begin{enumerate}

\item $s(\textbf{0})=0$ ($\textbf{0}$ is the null subspace).

\item $S(P^{\bot})=1-s(P)$

\item For any pairwise orthogonal
denumerable family of projections ${P_{j}}$ one has \newline
$s(\sum_{j}P_{j})=\sum_{j}s(P_{j})$
\end{enumerate}
For a study of the differences between classical and quantum
probabilities, see \cite{Gudder-StatisticalMethods} (chapter $2$).
Moreover, Gleason's theorem \cite{Gleason,Gleason-Dvurechenski-2009}
asserts that if $dim(\mathcal{H})\geq 3$, then the set of all
measures of the form (\ref{e:nonkolmogorov}) can be put into a
one-to-one correspondence with the set $\mathcal{C}$ of by all
positive, hermitian, and trace-class (normalized to unity) operators
in $\mathcal{B}(\mathcal{H})$. If $P\in\mathcal{P}(\mathcal{H})$,
the correspondence between $\rho\in\mathcal{C}$ and its induced
probability measure is given by

\begin{equation}\label{e:bornrule}
s_{\rho}(P)=\mbox{tr}(\rho P),
\end{equation}
\noindent where $\mbox{tr}(\cdots)$ stands for the trace operator in
$\mathcal{B}(\mathcal{H})$, i.e., the sum of all the eigenvalues.
Equation (\ref{e:bornrule}) is essentially Born's rule. Any
$\rho\in\mathcal{C}$ may be written as

\begin{equation}\label{e:convexity}
\rho=\sum_{i}p_{i}P_{\psi_{i}},
\end{equation}
\noindent where the $P_{\psi_{i}}$ are one dimensional projection
operators on the rays (subspaces of dimension one) generated by the
vectors $\psi_{i}$ and $\sum_{i}p_{i}=1$ ($p_{i}\geq 0$). Thus, it
is clear that $\mathcal{C}$ is a convex set. If the sum in
(\ref{e:convexity}) is finite, then $\rho$ is said to be of finite
range. Remark that in the infinite dimensional case the sum in
(\ref{e:convexity}) may be infinite in a non-trivial sense.
$\mathcal{C}$ is then a set of non-boolean probability
measures, closed by convexity, %(����Probar esto que digo en la
%siguiente frase!!!!!!)
that is also closed in the norm of $\mathcal{H}$. Let us remind the
reader that a {\sf lattice} is a partially ordered set (also called
a poset) in which any two elements have a unique supremum (the
elements' least upper bound; called their join) and an infimum
(greatest lower bound; called their meet). Lattices can also be
characterized as algebraic structures satisfying certain axiomatic
identities. Since the two definitions are equivalent, lattice theory
draws on both order theory and universal algebra. Semilattices
include lattices, which in turn include Heyting and Boolean
algebras. These "lattice-like" structures all admit order-theoretic
as well as algebraic descriptions \cite{lattice}.

\nd A very important notion for our purposes is that of
$\mathcal{L}_{\mathcal{C}}$, {\it the set of all convex subsets of
$\mathcal{C}$}. Any element of $\mathcal{L}_{\mathcal{C}}$ will be
itself a ``probability space", in the sense that it is a set of non
boolean probability measures closed under convex combinations (not
to be confused with the usual mathematical notion of sample space)
\cite{holiketal}. It can be shown that $\mathcal{L}_{\mathcal{C}}$
is endowed with a canonical lattice theoretical structure that can
be related to quantum entanglement and positive maps
\cite{holiketal}. The meet operation is given by set intersection,
the join by convex hull and the partial order by set inclusion. We
will use this lattice in order to express the Max-Ent protocol in a
different (but equivalent) form. It should be also clear that the
lattice operations mentioned above may be trivially defined in any
COM and this will be used in Section \ref{s:Generalization}.

\nd A general (pure) state can be written as

\begin{equation}
\rho=|\psi\rangle\langle\psi|,
\end{equation}
\noindent and we denote the set of all pure states by

\begin{equation}
P(\mathcal{C}):=\{\rho\in\mathcal{C}\,|\, \rho^{2}=\rho\}
\end{equation}
This set is in correspondence with the rays of $\mathcal{H}$ via the
association

\begin{equation}
\mathcal{F}:\mathbb{C}\mathbb{P}(\mathcal{H})\rightarrow \mathcal{C}
\quad|\quad [|\psi\rangle]\mapsto|\psi\rangle\langle\psi|,
\end{equation}

\noindent where $\mathbb{C}\mathbb{P}(\mathcal{H})$ is the
projective space of $\mathcal{H}$ and $[|\psi\rangle]$ is the class
defined by the vector $|\psi\rangle$
($|\varphi\rangle\sim|\psi\rangle\longleftrightarrow|\varphi\rangle=\lambda|\psi\rangle$,
$\lambda\neq 0$). If $M$ represents an observable, its mean value
$\langle M\rangle$ is given by

\begin{equation}\label{e:meanvalueoperator}
\mbox{tr}(\rho M)=\langle M\rangle
\end{equation}

Notice that the set of positive operators has the shape of a cone
while the set of trace class operators (of trace one) that of a
hiperplane. Thus, $\mathcal{C}$ is the intersection of a cone and a
hiperplane, embbeded in $\mathcal{A}$. Such structure (or
geometrical convex setting) is susceptible of considerable
generalization (see
\cite{Barnum-Wilce-2006,Barnum-Wilce-2009,Barnum-Wilce-2010} for an
excellent overview). In modeling probabilistic operational theories
one associates to any probabilistic system a triplet $(X,\Sigma,p)$,
where $\Sigma$ represents the set of states of the system, $X$ is
the set of possible measurement outcomes, and $p:X\times
\Sigma\mapsto [0,1]$ assigns, to {\it each outcome $x\in X$ and
state $s\in\Sigma$}, a probability $p(x,s)$ of $x$ to occur if the
system is in the state $s$. If we fix $s$ we obtain the mapping
$s\mapsto p(\cdots,s)$ from $\Sigma\rightarrow [0,1]^{X}$. We then
identify in this way {\it all the states of $\Sigma$ with maps of
such a form}. Focusing now attention upon their closed convex hull
we obtain the set $\Omega$ of possible probabilistic mixtures
(represented mathematically by convex combinations) of states in
$\Sigma$. One appreciates that we also obtain, for any outcome $x\in
X$, an affine evaluation-functional $f_{x}:\Omega\rightarrow [0,1]$,
given by $f_{x}(\alpha)=\alpha(x)$ for all $\alpha\in \Omega$. More
generally, any affine functional $f:\Omega\rightarrow [0,1]$ may be
regarded as representing a measurement outcome and thus use
$f(\alpha)$ to represent the probability for that outcome in state
$\alpha$. \vskip 3mm

\subsection{Effects}

\nd In the special case of quantum mechanics the set of all affine
functionals so-defined is called the set of effects. They form an
algebra (known as the \emph{effect algebra}) and represent
generalized measurements (unsharp, as opposed to sharp measures
defined by projection valued measures). Effect algebras have
important applications in the foundations of quantum mechanics
\cite{Muynck-POVM-2006} and in fuzzy probability theory
\cite{palmanova}. The specifical form of an effect in quantum
mechanics is as follows. A generalized observable or \emph{positive
operator valued measure} (POVM)
\cite{Busch-Lahti-2009,Thesis-Heinonen-2005,Ma-Effects} will be
represented by a mapping $E:\mathbf{B}(\mathbb{R})\rightarrow
\mathcal{B}(\mathcal{H})$ such that

\begin{enumerate}

\item $E(\mathbb{R})=\mathbf{1}$

\item $E(B)\geq 0,$, \,\,\mbox{for any}\, $B\in\mathbf{B}(\mathbb{R})$

\item $E(\cup_{j}(B_{j}))=\sum_{j}E(B_{j})$, \,{\rm for any
disjoint family} ${B_{j}}$

\end{enumerate}

\noindent The first condition means that $E$ is normalized to unity,
the second one that $E$ maps any Borel set B to a positive operator,
and the third one that $E$ is $\sigma$-additive with respect to the
weak operator topology. In this way, a POVM can be used to define a
family of affine functionals on the state space $\mathcal{C}$ (which
corresponds to $\Omega$ in the general probabilistic setting) of
quantum mechanics, as follows

\ben & E(B):\mathcal{C}\rightarrow [0,1] \cr &  \rho\mapsto
\mbox{tr}(E\rho) \label{probsett}.  \een
 Effects are then positive operators $E(B)$ which satisfy $0\leq E\leq\mathbf{1}$
 \cite{EffectAlgebras-Foulis-2001,Cattaneo-Gudder-1999}).
Let us denote by $\mathcal{E}(\mathcal{H})$ to the set of all
effects in quantum theory. We appeal to effects below in order to
define an special example of elements of
$\mathcal{L}_{\mathcal{C}}$, and also as a generalization of
conditions imposed in the Max-Ent protocol. Returning now to the
general model of probability states we may consider the convex set
$\Omega$ as the basis of a positive cone $V_{+}(\Omega)$ of the
linear space $V(\Omega)$. Thus, every affine linear functional can
be extended to a linear functional in $V(\Omega)^{\ast}$ (the dual
linear space). It can be shown that there is a unique unity
functional such that $u_{\Omega}(\alpha)=1$ for all
$\alpha\in\Omega$ (in quantum mechanics, this unit functional is the
trace function). Thus, the COM approach speaks of a triplet
$(A,A^{\sharp},u_{A})$, where $A$ is a  space endowed with a
strictly positive linear functional $u_{A}$ and $A^{\sharp}$ is a
weak-$\ast$ dense subspace of $A^{\ast}$, ordered by a chosen
regular cone $A_{+}^{\sharp}\subseteq A_{+}^{\ast}$ containing
$u_{A}$. Effects will be functionals $f$ in $A_{+}^{\sharp}$ such
that $f\leq u_{A}$.

\section{A geometrical expression for Max-Ent}\label{s:NewWay}

\nd Our idea is to re-discuss MaxEnt in terms of elements of
$\mathcal{L}_{\mathcal{C}}$ associated to conditions on given sets
of observables or effects. Consider the states

\begin{equation}
C_{(E,\lambda)}:=\{\rho\in \mathcal{C}\,|\,\mbox{tr}(\rho
E)=\lambda,\,\,\lambda\in[0,1]\}.
\end{equation}

\noindent $C_{E}$ is a convex set, and so, an element of
$\mathcal{L}_{\mathcal{C}}$. $C_{(E,\lambda)}$, represents all the
states for which the probability of having the effect $E$ is equal
to $\lambda$. Furthermore, there exists $\mathbb{S}$, a
$\mathbb{R}$-subspace of $\mathcal{A}$ (the set of bounded self
adjoint operators), such that

\begin{equation}\label{e:effectconvex}
C_{(E,\lambda)}=\mathbb{S}\cap\mathcal{C}
\end{equation}

\noindent and thus, $C_{E}$ is also an element of $\mathcal{L}$, the
lattice induced by the intersection of all closed subspaces of
$\mathcal{A}$ and $\mathcal{C}$
\cite{holiketal,Holik-Massri-Ciancaglini-2010,extendedql}. More
generally, if

\begin{equation}\label{e:setofmatrixes}
\langle R\rangle=r
\end{equation}

\noindent then the above equation may be considered as represented
by the set of density matrices which serve as a solution of it. The
ensuing set is obtained as the intersection of the Kernel of the
functional $F_{R}(\rho):=\mbox{tr}(R\rho)-r\mbox{tr}(\rho)$ and
$\mathcal{C}$. Accordingly, each equation of the form
(\ref{e:setofmatrixes}) (understood as an equation to be solved) can
be represented as an element $C\in\mathcal{L}$, and also as an
element of $\mathcal{L}_{\mathcal{C}}$. $C$ is also a closed set,
because it is the intersection of the kernel of a functional (which
is a closed subspace) and $\mathcal{C}$ \cite{holiketal}.

\noindent With such materials at hand, we can now re-express the
maximum entropy principle \cite{jaynes,katz} in  lattice theoretical
form. {\it Our point here is that  the set of conditions
(\ref{e:conditionsmean}) can be expressed in an explicit lattice
theoretical form as follows}. Using a similar procedure as in
(\ref{e:setofmatrixes}) we conclude that each of the equations in
(\ref{e:conditionsmean}) can be represented as a convex (and closed)
sets $C_{R_{i}}$. In this way we can now express conditions
(\ref{e:conditionsmean}) with the lattice theoretical expression

\begin{equation}
C_{max-ent}:=\bigcap_{i}C_{R_{i}}=\bigwedge_{i}C_{R_{i}}
\end{equation}

Now, $C_{max-ent}$ is also an element of $\mathcal{L}_{\mathcal{C}}$
(but not necessarily of $\mathcal{L}$) and we must maximize entropy
on $C_{max-ent}$. We have thus encountered a {\sf MaxEnt-lattice
theoretical expression}: {\it given a set of conditions represented
generally by convex subsets $C_{i}$, one should maximize the entropy
in the set $C_{max-ent}=\bigwedge_{i}C_{i}$.}

\section{Generalization}\label{s:Generalization}

A (generalized) observable in a test space $\Omega$ will be given by
a function $F:x\mapsto F_{x}$ from subsets of a given outcome set
$X$ into $A(\Omega)$ satisfying

\begin{enumerate}

\item $F_{x}\geq \mathbf{0}$

\item $\sum_{x\in E}F_{x}=u$

\end{enumerate}

This is the natural extension of the notion of POVM to any COM
(extension which includes the more restricted case of projection
operator valued measures for the quantum case). If
$\omega\in\Omega$, we can construct the probability function
$p_{\omega,F}(x)=F_{x}(\omega)$, and this pulls back a map
$F^{\ast}:\Omega\rightarrow\Delta(E)$, with
$F^{\ast}(\omega)=p_{\omega,F}$. With this at hand, we can easily
generalize probabilities (and so, when they appear, mean values) as
linear conditions. With this procedure we will obtain convex sets,
and proceed similarly as in Section \ref{s:NewWay}. For example,
suppose that the outcomes of our observable $F$ are represented by
the discrete set of real numbers $\{f_{i}\}_{i\in\mathbb{N}}$. Then
we can write the mean value of $F$ in the state $\omega$ as:

\begin{equation}\label{e:meanvalueCOM}
\langle F\rangle_{\omega}=\sum_{i}f_{i}F_{f_{i}}(\omega)
\end{equation}

\noindent The probability functions $F_{f_{i}}(\cdot)$ are linear
operators defined on $A$ taking values in the interval $[0,1]$ when
restricted to states. As they are linear, we can define the
functional

\begin{equation}\label{e:meanvalueCOM2}
\langle F\rangle:A\rightarrow\mathbb{R}\,\,|\,\, \langle
F\rangle(a)\mapsto\sum_{i}f_{i}F_{f_{i}}(a)
\end{equation}

\noindent If we now consider the new functional

\begin{equation}
\langle F\rangle_{r}:A\rightarrow\mathbb{R}\,\,|\,\, \langle
F\rangle_{r}(a)\mapsto\sum_{i}f_{i}F_{f_{i}}(a)-ru(a)
\end{equation}

\noindent

\noindent then, the Kernel of $\langle F\rangle_{r}$ will be a
linear subspace of $A$, call it $\mathbb{S}_{r,F}$. If we now
restrict to the elements of $\Omega$ (and any $\omega\in\Omega$
satisfies $u(\omega)=1$), we will have that solutions of the
Equation \ref{e:meanvalueCOM} can be represented with the convex set

\begin{equation}
C_{r,F}:=\mathbb{S}_{r,F}\cap\Omega
\end{equation}

\noindent More generally, even if our measurement outcomes are just
labeled by an index set and we are not interested on mean values but
only on probabilities (as is generally the case for many POVM's),
conditions will have the form

\begin{equation}\label{e:probaeffect}
p_{\omega,F}(x)=F_{x}(\omega)=\alpha
\end{equation}

\noindent and following an analogous reasoning line as above, it can
also be put in the form $\mathbb{S}_{\alpha,F}\cap\Omega$, with
$\mathbb{S}_{\alpha,F}$ a subspace which depends on the real number
$\alpha$ and the observable $F$.

\noindent In any case, the construction given above provides the
path for generalization of the procedure outlined in Section
\ref{s:NewWay} to any COM. If $E_{i}$ represent observables and the
brackets $\langle\cdot\rangle$ are meant to express either
conditions of the form of Equations \ref{e:meanvalueCOM} or
\ref{e:probaeffect}, suppose as given a set of conditions of the
form

\begin{eqnarray}\label{e:conditionsmean2}
\langle E_{i}\rangle=e_{i};\,\, {\rm for \,\,all\,\,} i.
\end{eqnarray}

\noindent Then, because of the above discussion, the set of
conditions \ref{e:conditionsmean2} may be expressed as a collection
of convex subsets of the form
$C_{e_{i}}=\mathbb{S}_{e_{i}}\cap\Omega$. And the Jaynes method
tells us that we must maximize entropy on the convex set:

\begin{equation}
C_{MaxEnt}=\bigwedge_{i}(\mathbb{S}_{e_{i}}\cap\Omega)
\end{equation}

\noindent It is important to remark here that our construction does
not only restricts to projective measurements, but it includes the
possibility of imposing conditions on POVMs. Thus, we provide here a
generalization of the Max Ent protocol for generalized measurements.

\noindent We notice that the specific form of the entropy to
maximize depends on the specifical COM which is being used. For
example, for the particular cases of quantum theory and a classical
model, von Newmann and Shannon entropies are used respectively in a
natural way. The election of the particular function to maximize on
$C_{MaxEnt}$, does not depends on the general formulation but in the
specifical characteristics of the model. An interesting question
would be to set up general conditions for entropies in any model,
but the existence of such conditions is hard to believe, because of
the proliferation of entropies and contexts.

\section{Conclusions}\label{s:Conclusions}

In this work we reviewed the celebrated E. T. Jaynes max-ent
principle. After that, we studied generalizations of probabilistic
models, the COM approach. In Section \ref{s:NewWay} we gave a new
expression for max-ent using a lattice theoretical fashion,
expressing conditions as convex sets and taking the maximum in their
set theoretical intersection, i.e., lattice conjunction. In Section
\ref{s:Generalization}, we used the key fact that the probability
function of COM preserves convexity in order to link the COM
approach with the results of Section \ref{s:NewWay}, and gave a
generalization for max-ent. This generalization also includes a
max-ent protocol for conditions on POVMs in any COM. This kind of
generalizations may be useful to extend the powerful max-ent
approach for more general theories and also yields light into its
mathematical and geometrical structure/background.

\vskip1truecm

\noindent {\bf Acknowledgements} \noindent This work was partially
supported by the following grants: .

\end{document}